# Small-Angle Neutron Scattering study of solubilization of tributyl phosphate in aqueous solutions of L64 Pluronic triblock copolymers

J. CAUSSE [§, ¢], J. OBERDISSE [±], J. JESTIN [¥], S. LAGERGE [§, *]

[§] Institut Charles Gerhardt CNRS UMR-5253 équipe « Agrégats Interfaces Matériaux pour l'Energie », Université Montpellier II, Case 015, Place E. Bataillon, 34095 Montpellier Cedex 05, France.

[¥] Laboratoire Léon Brillouin (LLB, UMR 12 CEA-CNRS), CEA Saclay, 91191 Gif-sur-Yvette Cedex, France

[±] Laboratoire des Colloïdes, Verres et Nanomatériaux (LCVN, CNRS UMR-5587), Université Montpellier II, Case 015, Place E. Bataillon, 34095 Montpellier Cedex 05, France.

[¢] Laboratoire des Procédés Avancés de Décontamination (DEN/DTCD/SPDE/LPAD), CEA Marcoule, Bât. 438, BP17171, 30207 Bagnols Sur Cèze, France

* To whom correspondence should be addressed.

Tel.: +33 (0)4 67 14 46 20. Fax: +33 (0)4 6714 33 04. E-mail: slagerge@univ-montp2.fr

## ABSTRACT

We have studied the solubilization behaviour of tributylphosphate (TBP) in aqueous solutions of L64-Pluronics, using light and small angle neutron scattering (SANS). Varying the temperature and the oil-content, the system presents a non trivial phase behaviour. In particular, at 308K, a first solubilization followed by an emulsification failure and a resolubilization is found. We have measured the microstructure by SANS and characterized the microemulsion droplet core-size, corona-thickness, polydispersity, and interactions. It is shown that at low oil content, the system is made of small swollen micelles. After the phase separation, the resolubilization is carried by larger oil droplets decorated by copolymer. From specific surface measurements at large angles, a surprising change in surfactant conformation is found to accompany this morphological evolution which is also supported by previous results obtained from 1H NMR experiments. In independent measurements, our structural modelling is confirmed using contrast-variation SANS.

**INTRODUCTION**

The solubilisation of polar organic additives in aqueous solutions containing amphiphilic block copolymers is one of the main fundamental aspects of the physical chemistry of microemulsions. It is also of wide practical importance, e.g. for nuclear decontamination, separation of non-ferrous metals, rare-earth metals, and actinides [1]. The solubility of such contaminants depends on the self-assembled structure of the amphiphiles in water, i.e. on physico-chemical characteristics like the hydrophilic-hydrophobic balance of the copolymer, or the spontaneous curvature of the hydrophilic-hydrophobic interface. In this article, the structure of triblock copolymer aggregates in water is studied with a polar oil used in nuclear decontamination (tributylphosphate, TBP) as a prototype contaminant. This lipophilic compound is used as a complexant in liquid-liquid extraction cycles of radioactive metals such as, for example, uranium, plutonium and rhodium [2-4].

The triblock copolymer studied here is Pluronic L64, a well-known amphiphilic copolymer consisting of two hydrophilic side-blocks (poly(ethylene oxide), PEO) and an hydrophobic core block (poly(propylene oxide), PPO). In water, it forms micelles above a given concentration (cmc) and temperature (critical micelle temperature, cmt) [5-11]. Pluronics are used as emulsifiers, solubilizers, foaming/defoaming agents, wetting agents, or vector agents in controlled release [10]. Their aggregation and solubilisation properties have been widely investigated in the past decade [12-22]. One major advantage of Pluronics is that their amphiphilic properties may be controlled not only by their chemical structure (blocks lengths) but also by the temperature [9, 23]. PPO and PEO polarity in aqueous solution decreases as the temperature rises leading to dehydration of the PEO chains. The solubility of these copolymers thus decreases and the cmc is lower at higher temperatures. At the Cloud Point Temperature (CPT), all PEO chains are dehydrated, resulting in the precipitation of the polymeric chains.

Block copolymers with various lengths of hydrophobic and hydrophilic blocks are promising molecules for solubilisation due to their simultaneous amphiphilic (adsorption at interfaces) and polymeric (stabilization of colloidal systems) behaviour. Because of their marked temperature-dependant amphiphilic character, Pluronics micelles are capable of solubilising organic molecules of different polarities [28]. In such studies, the determination of the maximum additive concentration is an important feature [29, 30]. Many studies reported solubility measurements and partition coefficient of the host molecule between water and the aggregates, using UV-spectroscopy [31], HPLC technique [32] and fluorescence spectroscopy [33]. They usually focussed on hydrophobic compounds such as drugs [31, 34], or benzene derivatives [11, 35-38], showing that aqueous solutions of Pluronics significantly enhanced the solubility of hydrophobic compounds. Others also evidenced the enhancement of the copolymer aggregation number through the incorporation of o-xylene [39, 40] or chlorinated additives [41] in Pluronic aggregates.

The location of the solute in the micelle varies with the nature of the amphiphile, especially with polar oils with intermediate properties. King et al studied the solubilisation of gases such as methane, ethane or propane and showed that these species were located close to the hydrophobic core of the Pluronics aggregate [42]. Vauthey et al. showed that slightly hydrophobic polar molecules were located in the aggregate core for low concentrations and in the micellar palisade for higher concentrations [43]. The location of the solubilized species may be obtained from [1]H NMR through the dependence of chemical shifts or linewidths on concentrations [44, 45-49]. Other techniques such as Self Diffusion [50, 51], Relaxation Time measurements [52-54], fluorescence spectroscopy and time resolved fluorescence [55-58] have also provided indications on the solubilisation site of molecules in Pluronic aggregates. In spite of this large body of work, the solubilisation in Pluronics aggregates still remains under debate, with scarce literature on the solubilisation of polar oils and their location within the aggregates.

In the present paper, the microstructure of TBP-Pluronic aggregates is investigated by Small Angle Neutron Scattering (SANS) as a function of the TBP concentration [61, 62]. SANS is a suitable technique to characterise microemulsions [78-81], and phases of block copolymers [19, 63-66, 41, 67, 68], in particular due to deuteration methods. Structural changes caused by the addition of the oil in micelles have been evidenced [41, 69, 70]. Möller et al. studied the solubilization of butanol in alkyl glucosides micelles [69]. They found the site of solubilization of this very polar oil close to the palisade of the micelle. Later De Lisi et al. demonstrated the ability of aqueous block copolymer – surfactant mixtures in solubilizing chlorinated organic compounds [70].

After the experimental section of this article, the solubility curve is discussed in the phase diagram section. Samples of increasing oil content have been studied by DLS, and more in detail by SANS. The droplet growth has been followed in this ternary oil-in-water microemulsion systems differing only by the TBP content. In a second experiment, we have used contrast variation by using fully deuterated TBP [72, 73], giving access to the distribution of the TBP molecules within the aggregates.

**EXPERIMENTAL SECTION**

**Material and sample preparation.** The ternary microemulsion systems are made of aqueous solutions of amphiphilic triblock copolymers including the polar oil, TriButylPhosphate (TBP). A Pluronic L64 copolymer (poly(ethyleneoxide)–poly(propyleneoxide)–poly(ethyleneoxide)), $(EO)_n–(PO)_m–(EO)_n$, n = 13, m = 30) was obtained as a gift from BASF Corporation (France). This polymeric material is known to contain small contaminating quantities of hydrophobic impurities [84] which were removed following a purification procedure by dissolution/ precipitation using hexane (purity exceeding 99 %) [88] reported in previous works [61, 62]. The water used in these experiments was distilled and deionised with a Millipore "Super Q" system (18 M$\Omega$ cm). Deuterated water ($D_2O$, 99.9 % deuterated) used in SANS experiments was purchased from Sigma. TBP, ($O=P(–O–(CH_2)_3–CH_3)_3$ with purity $\approx$ 99 % was supplied by Merck (France). Deuterated TBP provided by Eurisotop (France) was used for the solvent contrast variation experiments in SANS. TBP is a complexing agent which is widely used in the selective extraction of U and Pu from organic solutions [3, 4]. It is a polar oil ($\varepsilon$ = 8.09 and $\mu$ = 3.1 Debye [1]) which is soluble in deionised water to a very small extent ; its saturation concentration in pure deionised water, $c_{sat}$, is 1.6 x $10^{-3}$ mol kg$^{-1}$ (= 0.42 g/l), value which we use to express the oil concentration as $c/c_{sat}$. TBP molecules in aqueous solution at concentration close to the saturation value exhibit surface active properties ($\gamma$ = 42.1 mN m$^{-1}$) owing to their polarity.

Polymeric solutions were prepared by dissolving a given mass of L64 in deionised water previously filtered through a MiniSart 0.22 $\mu$m cellulose acetate filter. Once the desired concentration of amphiphilic polymer was obtained, the sample was maintained under gentle agitation for one night at constant temperature. Then the solutions were kept under relaxation for one day before performing all the analysis. The same procedure was used in $H_2O$ and $D_2O$.

**Small-angle neutron scattering (SANS).** Experiments were performed at the Laboratoire Léon Brillouin (LLB), CEA Saclay (France), on beamline PACE. The scattered intensity has been measured over a wide range of scattering vectors q, between 0.005 Å$^{-1}$ and 0.4 Å$^{-1}$, using three configurations (D = 1.07 m λ= 5 Å ; D = 2.87 m  λ = 6 Å ; D = 4.57 m λ = 12 Å, with Δλ/λ = 10 % ) where D and λ are the sample-to-detector distance and the neutron wavelength, respectively. The scattered intensity was corrected for empty cell scattering and incoherent background, and put on an absolute scale according to standard procedures [74]. The resolution function was taken into account in our modelling following the literature [82, 83].

For all experiments, flat quartz cells with optical path length of 1 or 2 mm were used according to the hydrogenation degree of the solutions. Aqueous solutions of copolymer with or without added TBP were loaded into the sample cells at room temperature, then placed in the instrument sample chamber at the desired temperature (308 K) and allowed to equilibrate for several hours before data were taken. The temperature of the sample was maintained by a circulating water bath. All our intensities have been measured in absolute units ($cm^{-1}$), and information on aggregate mass can be extracted knowing the average scattering length densities of the constituents: ρ =$0.43 \times 10^{10}$ $cm^{-2}$ (L64), $0.16 \times 10^{10}$ $cm^{-2}$ (TBP), $5.78 \times 10^{10}$ $cm^{-2}$ (TBPd). Contrasts Δρ with respect to the solvent can then be calculated. The steps of the fitting procedures are outlined in the appendix.

**Dynamic light scattering (DLS).** Hydrodynamic radii were measured by Photon Correlation Spectroscopy with a Zetasizer 3000 Hsa apparatus (Malvern Instrument Ltd.) which operates with a 10mW He-Ne laser at λ = 633 nm. Temperature was controlled at 308K using a Peltier element. The Contin algorithm was used to fit the experimental data and to obtain the diffusion coefficient D of the aggregates, which gives the hydrodynamic radius of the droplets using the Stokes-Einstein relationship. Before analysis, the sample solutions were filtered with a 0.1 μm Millipore membrane into glass cuvettes of optical quality. The cinematic viscosity (ν) of the polymeric solutions and the ternary systems (TBP-water-L64) were evaluated by means of a capillary viscosimeter (Ubbelohde). An Ubbelohde tube (Viscosimetric MS, Fica) with diameter of 0.47 mm (K = $0.00321$ $mm^2$ $s^{-2}$) was employed. The corresponding densities (ρ) were measured with a PAAR DMA 602 density meter. So we could finally determine the dynamic viscosities (η = ν x ρ) of the solutions which are involved in the Stokes-Einstein relationship.

**Turbidity measurements** were made using a spectroscopic technique (Metrohm 662 photometer at 600 nm) in order to detect the solubility limit of TBP in the polymeric solution. The solubilization procedure was carried out as follows. Small aliquots (20 µl) of the pure TBP solution was injected stepwise, using an external syringe, into a beaker containing 25 ml of an homogeneous solution of Pluronics above the aggregation concentration. After each injection, the system was kept under agitation during 15 minutes before measuring the turbidity change. It is thus possible to detect the turbidity changes associated with subsequent steps, i.e. occurring upon introduction of TBP in the polymeric solutions and to follow the phase behaviour of the systems (solubilisation process) step by step. The turbidity change after each injection was followed by measuring the voltages output U of the spectrode, which directly reflects the light transmission through the solution. Thus solubilization curves could be constructed by plotting the evolution of voltage against the concentration of TBP in the medium, U = f(C).

## RESULTS AND DISCUSSION

**Phase diagram.** The solubility of TBP in aqueous solutions of the amphiphilic copolymer L64 at 10 wt% has been previously studied in details using turbidity measurements at T = 308 K [61, 62]. The results are shown in Figure 1a, where the normalized voltage $U/U_r$ is plotted as a function of the normalized TBP concentration $c/c_{sat}$. $c_{sat}$ is the solubility of TBP in pure deionised water. $U_r$ is the reference voltage obtained in the pure polymeric solution, U the voltage obtained for a L64 solution at a given TBP content, and $U/U_r$ is thus directly proportional to the light transmission through the sample.

The maximal transmittance of the solution is obtained with monophasic systems. Consequently the extent of the plateau region at $U/U_r \approx 1$ gives direct information on the solubilization capacity of the L64 solution. Surprisingly, the system exhibits a first phase separation between $c/c_{sat}$ = 70 and 150, and becomes clear and monophasic again at higher oil content ($c/c_{sat}$ > 150). This phase seems to contain slightly bigger objects with a higher capacity to scatter light, causing an appreciable decrease in transmission. The typical evolution of the system reported in Figure 1a has been widely discussed in a previous paper [61]. The system exhibits two monophasic regions separated by a two phase domain (70 < $c/c_{sat}$ 160). We could conclude that the phase separation observed at $c/c_{sat}$ = 70 and 280 are due to the precipitation of the copolymer and foration of large emulsion droplets, respectively.

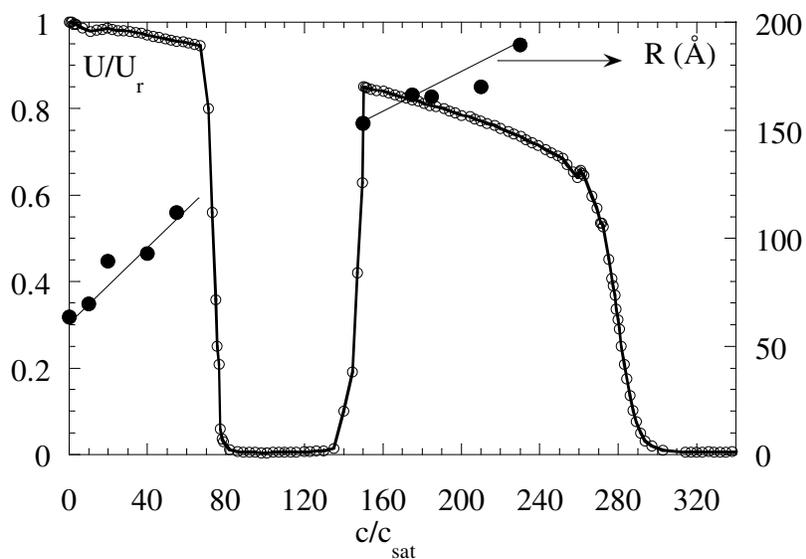

**Figure 1a:** Phase behaviour of a 10 wt% aqueous solution of L64 as a function of TBP concentration, at 308 K, as explored by turbidity measurements ($\circ$). Starting from the oil-free system ($c/c_{sat} = 0$), the system is first monophasic, then displays a phase transition around $c/c_{sat}$= 70 and becomes very turbid. At higher oil concentration, $c/c_{sat}$= 150, a second transition towards a new transparent phase is observed. The hydrodynamic radius is also plotted ($\bullet$).

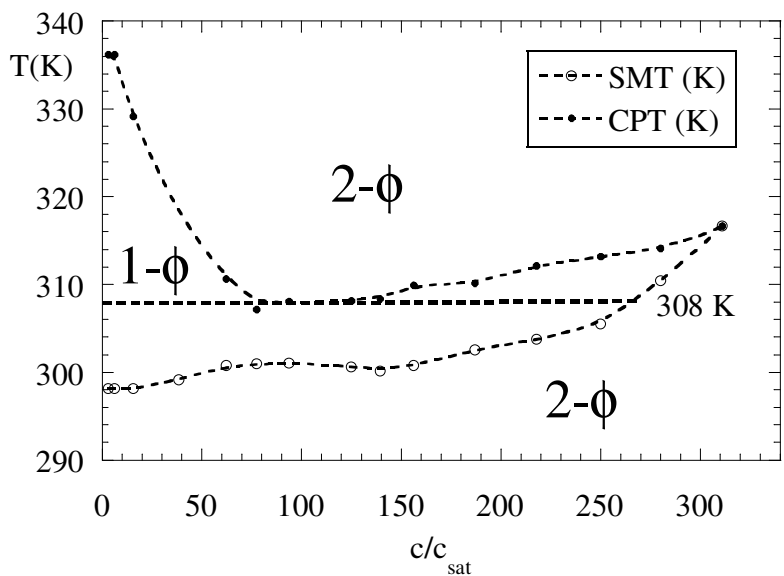

**Figure 1b:** Cut through the phase diagram of the three-component system (L64 at 10 wt% – TBP – water). This diagram shows the evolution of both the cloud point temperature (CPT, black symbols) and the solubilization minimal temperature (SMT, open symbols) against the normalized TBP concentration $c/c_{sat}$. 1-$\Phi$ and 2-$\Phi$ denote one- and two-phase regions, respectively.

On the same graph in Figure 1a we have reported the evolution of the average hydrodynamic radius of the aggregates (black circles) measured using DLS. The smallest objects, at $c/c_{sat} = 0$, have the typical size of small polymeric micelles. This suggests the presence of microemulsion droplets, which grow as the oil content is increased, up to an emulsification failure at $c/c_{sat} = 70$. Surprisingly, at even higher oil concentration, the TBP is solubilised again, in the form of bigger microemulsion droplets approaching apparent hydrodynamic radii of 200 Å. Due to lower transmission, these data may be affected by multiple scattering effects. It turns out, however, that they will be fully confirmed by the SANS analysis. Note that the presence of microemulsion droplets is in agreement with [1]H NMR analyses of various three components systems (Pluronic – TBP - water) [62] which show that TBP molecules are preferably inserted into the hydrophobic core of the Pluronic micelles. The [1]H NMR analysis also evidenced an evolution of the hydration state of the hydrophobic core following addition of TBP in the micellar solutions. During the addition of TBP, the microemulsion structure seems to turn from spherical micelles into nanodroplets of pure TBP stabilized by the Pluronic.

The emulsification failure is directly related to the spontaneous curvature of the hydrophobic-hydrophilic interface, which is itself governed in non ionic surfactant systems by the temperature, due to varying PEO-headgroup hydration. It seemed thus natural to explore the phase diagram for different temperatures using the turbidity as a probe. The resulting cut at fixed concentrations of L64 through the phase diagram, as obtained from previous turbidity data [61], is displayed in Figure 1b.

In Figure 1b, a one phase region (denoted 1-Φ) extending up to rather high TBP concentrations can be found at intermediate temperatures. It is limited at high T by the Cloud Point Temperature commonly defined for the binary systems of water and non ionic amphiphiles. Here it depends on the amount of oil, which indicates that the conformation of the PEO-groups of the surfactant molecules evolves, suggesting morphological changes. The low-T phase boundary represents the Solubilization Minimal Temperature, i.e. the lowest temperature required to obtain the suitable hydrophobicity of Pluronic micelles allowing TBP solubilisation at a given concentration. It becomes clear from the phase diagram that along the chosen isotherm of 308 K, the system is first monophasic, and then hits the two-phase region (denoted 2-Φ) due to the strong decrease of the CPT with added oil, before solubilising oil again at higher oil concentrations. This curious behaviour motivated the present study of the microstructure using small angle neutron scattering.

**Structure of pure L64 in $D_2O$:** Before characterizing the morphologies in presence of TBP, we have performed a structural study by Small Angle Neutron Scattering (SANS) of the pure aqueous copolymer solutions at 308 K. As one can see from the phase diagram in Fig. 1b, a transparent phase of pure copolymer L64 is formed in the absence of TBP. We have characterized the morphology of the aggregates by SANS in $D_2O$ for obvious contrast reasons, the average scattering length density of L64 being $0.43 \times 10^{10}$ cm$^{-2}$. The results are shown in Figure 2 for three different concentrations (0.2%wt, 5%wt, 10%wt), after subtraction of the unimer scattering for the two higher concentrations as explained below. At the highest dilution, which is below the cmc ($\approx$0.35%wt, cf. supporting information SI) the scattering pattern I(q) is due to the individual unimers in solution. The low-q limit $I_o$ (0.023 cm$^{-1}$) corresponds to a dry volume of 3300 Å$^3$, and the radius of gyration, deduced from the Guinier fit - eq.(1) applied to unimers - in Fig.2, is 18.5 Å. Both values clearly correspond to individual molecules, which have a dry volume of 4640 Å$^3$ estimated from the molecular mass. The discrepancy may be due to the low scattered intensity in this case of scattering by individual molecules, but the order of magnitude remains correct.

Above the cmc, we have measured the structure at two different concentrations. At 5%wt (4.8%v), the intensity is considerably higher than what would be expected from single molecules at the same concentration, which illustrates that aggregation has taken place. The aggregates are sufficiently dilute, and no interaction peak is visible in the intensity. We have based our analysis on the aggregation number of L64 at 35°C – about 35 – , which is known from the literature [20]. This number leads to a strong overestimation of $I_o$, which can – in absence of interaction - only be due to the coexistence of smaller objects, i.e. (non micellized) unimers, with micelles. This behaviour is consistent with our surface tension measurements (cf. SI), and has been also reported in the literature [9]. Such a coexistence in solution corresponds to a majority of 3.67%v of copolymer as unimers, and only the remaining $\Phi_{mic}$=1.13%v as micelles of aggregation number 35. In order to subtract the signal of the unimers, we have thus rescaled the unimer intensity measured at 0.2%wt to its value in solution, and deduced it from the measured 5%wt intensity. The result is shown in Fig. 2. These intensities serve as starting point for the following discussion in presence of TBP-oil using the Guinier approximation:

$$I = I_0 \exp(-q^2 R_g^2/3) \tag{1}$$

where $R_g$ is the radius of gyration of the finite-sized aggregates. The simplest Guinier fit parameter $I_o$ gives a typical dry volume of micelles of $162.5 \times 10^3$ Å$^3$, which corresponds to a dry radius of 33.8 Å. The q-dependence, however, corresponds to a Guinier radius of a homogeneous sphere of 45.7 Å deduced from $R_g$ (= 35.4 Å). The larger spatial dimension of the micelle suggests that they are swollen, which is due to the hydrated PEO layer. This motivated further fitting using a core-shell model, and acceptable fits were obtained with a shell thickness $\Delta$ = 35 Å and a pure PPO core radius of $R_c$ = 28.9 Å, with a log-normal polydispersity of $\sigma$ = 15% on the core radius. This fit could be further improved by assuming that some PPO (10%) participated in the highly hydrated shell containing 90% $D_2O$. It is shown in Fig.2. The total radius of the micelle, about 65 Å, is consistent with the DLS-results which gave hydrodynamic radii of about 63.5 Å for 10%wt-samples (cf. Fig.1a).

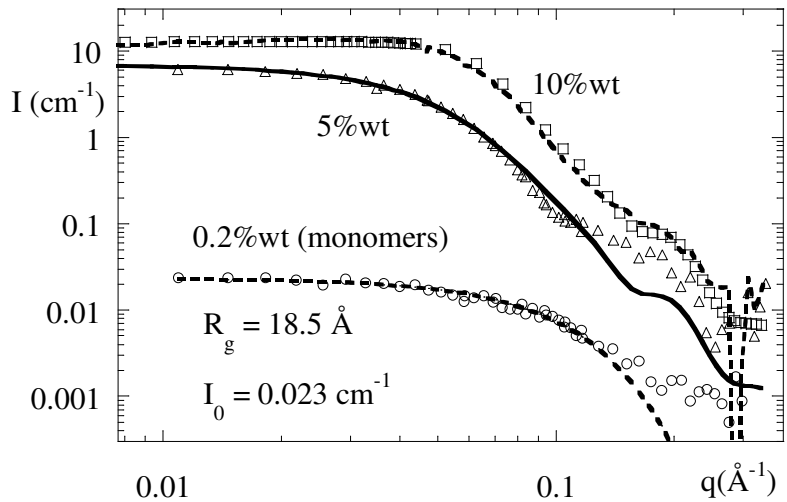

**Figure 2:** Scattered intensity of solutions of Pluronics L64 in $D_2O$ at three concentrations (0.2%wt, 5%wt, 10%wt), compared to model calculations, after subtraction of unimer scattering. A Guinier fit is shown for 0.2%wt, and polydisperse core-shell micelle models for the higher concentrations. $N_{L64}$ = 35, $R_c$ = 27.9 Å, $\Delta$ = 35 Å, $\sigma$ = 15%, 10% PPO in shell (5%wt), and $N_{L64}$ =35, $R_c$ = 27.9 Å, $\Delta$ = 35 Å, $\sigma$ = 15%, 10% PPO in shell (10%wt), the latter in combination with a PY-structure factor ($\Phi_{HS}$ =14.5%, $R_{HS}$ =45Å).

At the highest concentration 10%wt, finally, which is also shown in Fig. 2, an interaction peak is visible around 0.04 Å$^{-1}$. Again the free unimer scattering has been subtracted, supposing a constant unimer volume fraction of 3.67%v. This leaves a remaining volume fraction of micellized copolymer of $\Phi_{mic}$ = 6.0%v. We treat this case with some detail, as it will serve as an example of what follows. The low-q intensity is about 11 cm$^{-1}$, whereas 6%v of these micelles would scatter up to 35.1 cm$^{-1}$ in pure form factor scattering. This reduction to 31.3% is due to the weaker isothermal compressibility, i.e. it corresponds to S(0) = 0.313. As all aggregates are supposed to be spherical and uncharged, the structure factor S(q) can be described with the Percus-Yevick structure factor for hard spheres [75, 76]. Two parameters are needed, the hard sphere volume fraction $\Phi_{HS}$, and the interaction radius $R_{HS}$. $S(0, \Phi_{HS})$ follows from a straightforward calculation [74, 77], and we find $\Phi_{HS} = 14.5\%$ to satisfy this condition. This implies that the micelles occupy more space than one would guess from their dry radius of 33.8 Å, and using volume conservation, $R_{HS} = 45.4$ Å. It is interesting to note that this is the same size as the one found from the Guinier analysis of the more dilute sample. In Fig. 2, a combination of the hard-sphere structure factor, multiplied by the same core-shell form factor of polydisperse spheres as with the 5% sample (and again convoluted with the resolution function), is shown to reproduce the data well up to intermediate q-values. This was to be expected, because plotting the reduced I(q)/$\Phi_{mic}$ for 5% and 10%wt gave a nice superposition at intermediate angles, the influence of the structure factor being visible only at low q.

To summarize, the pure L64-micelles at 35°C are well described by a polydisperse core-shell model ($N_{agg}$ = 35, $R_{core}$ = 27.9 Å, polydispersity $\sigma$ = 15%, $\Delta$ = 35 Å), if one takes repulsive hard-sphere interactions into account at higher concentrations. It is underlined that we have followed the literature value of $N_{agg}$ = 35 [20], which implies that the observed level of intensity can only be explained by a substantial amount of free unimers in solution. Such a progressive micellisation is compatible with the surface tension data (cf. SI). As the intensity of oil-containing solutions strongly dominates possible unimer contributions, this discussion has no impact on what follows.

**Structure in presence of TBP, before phase separation:** The evolution of the scattering patterns of the three component system L64 (10%wt)-TBP-D$_2$O with increasing amount of TBP in the medium (c/c$_{sat}$ = 0, 25, 50, all before the phase separation) is shown in Fig. 3. We have again subtracted the unimer contribution, although it is almost completely negligible here. The low-q intensity is found to increase with increasing oil content, and all curves superimpose nicely in the so-called Porod-regime. In this high-q regime indicating well-defined interfaces, the scattering is proportional to q$^{-4}$, the prefactor A = 1.3 10$^{28}$ cm$^{-5}$ being directly related to the specific surface S/V [74] :

$$A = 2\pi \, \Delta\rho^2 \, S/V \qquad (2)$$

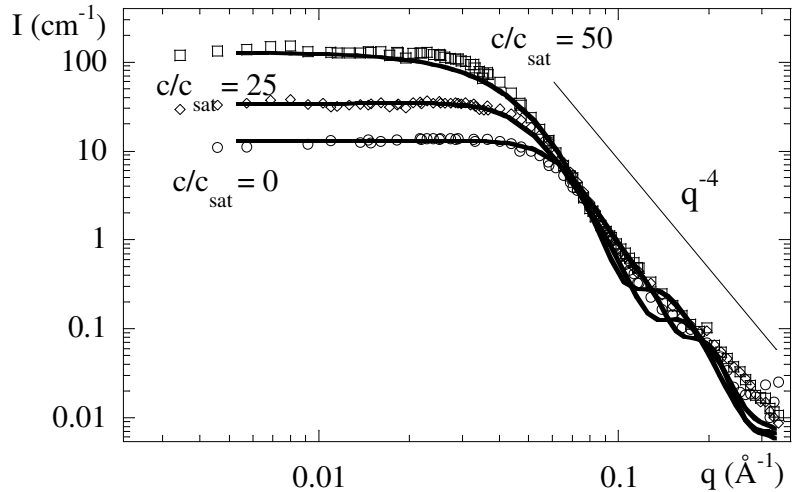

**Figure 3:** Scattered intensity of solutions of Pluronic L64 and TBP before the phase separation, at various concentrations in $D_2O$. The mass fraction of Pluronics is fixed to 10%wt, with three concentrations of TBP ($c/c_{sat}$ = 0; 25, 50). The curves are compared to core-shell model calculations (**0:** $N_{L64}$ = 35, $R_c$ = 27.9 Å, $\Delta$ = 35 Å, $\sigma$ = 15%, 10% PPO in shell; **25:** $N_{L64}$ = 49, $N_{TBP}$ = 88, $R_c$ = 35.1Å, $\Delta$ = 35 Å; **50:** $N_{L64}$ = 65, $N_{TBP}$ = 233, $R_c$ = 41.2 Å, $\Delta$ = 35 Å) combined with a PY-structure factor (see text for details).

The specific surface is the same in the three cases, as is the polymer volume fraction (6%). This suggests that the surfactant covers the hydrophobic-hydrophilic interface, as one might expect. From the ratio of specific surface and number density of L64 molecules, a headgroup area of 215 Å$^2$ per PEO group can be determined. Such values are compatible with those reported in the literature for similar copolymers, e.g. P123 with about 200 Å$^2$ per PEO group [71]. This head group area is also in good qualitative agreement with the one found by surface tension measurements of dilute solutions of pure L64 (246 Å$^2$), as shown in the SI. Assuming spherical objects of average contrast (polymer and oil in $D_2O$) of $\Delta\rho$ = 6.1 $10^{10}$ cm$^{-2}$, we find the same specific surface of S/V = 5.56 $10^5$ cm$^{-1}$ = 3 $\Phi_{mic}$/R$_{Porod}$ for all samples, where R$_{Porod}$ is the Porod radius, and the TBP is included in the total micellized volume fraction $\Phi_{mic}$. This specific surface is compatible with objects of Porod radius 32.4 Å ($c/c_{sat}$ = 0, total micellized volume fraction $\Phi_{mic}$ =6.0%), 38.3 Å ($c/c_{sat}$ =25, $\Phi_{mic}$ =7.1%), and 44.2 Å ($c/c_{sat}$ = 50, $\Phi_{mic}$ = 8.2%). In absence of oil, this radius is close to the core-radius of the pure L64, and suggests that this interface dominates the Porod scattering. It is possible to relate the increase in radius to an increase in volume of each micelle by a factor of 1.65 (resp. 2.54) for $c/c_{sat}$ =25 (resp. 50), which is caused by the incorporation of TBP-oil in the hydrophobic part of the micelles. Simultaneously, the number of micelles per unit volume must decrease approximately by 50%, because otherwise the specific surface could not be maintained constant upon simultaneous increase of the volume fraction. If we assume a homogeneous distribution of L64 and TBP in all micelles, the L64-aggregation number increases from 35 to 49 and finally 65, whereas the number of TBP-molecules per micelles increases from 0 to 88, and finally 233 at $c/c_{sat} = 50$.

A more precise analysis using a core-shell model together with the Percus-Yevick structure factor has lead to satisfying fits for these three samples. The results are also shown in Fig. 3, and the fitting parameters are given in the caption. The underlying model assumes that the TBP is solubilised in the micellar core. For a given number of TBP molecules per micelle, the number of micelles per unit volume can be directly deduced from mass or volume conservation (molecular volumes $V_{L64} = 4640 \ \text{Å}^3$, $V_{TBP} = 455 \ \text{Å}^3$). The scattering function has been calculated for a given geometry, i.e. core radius $R_c$, its polydispersity $\sigma$, and shell thickness $\Delta$, with a coherent set of parameters (contrasts, masses). Note, e.g., that for given aggregation numbers from the above Porod analysis, the average core radius $R_c$ is fixed by the volume of the hydrophobic parts, and only shell thickness $\Delta$ (and polydispersity $\sigma$, the main effect of which is smoothing of oscillations at large q) can be adjusted. The best fits show that the shell stays highly hydrated (>90% of $D_2O$, $\Delta = 35 \ \text{Å}$), and that the core region is mainly responsible for the scattering, due to its high contrast with the solvent. At $c/c_{sat} = 25$, the core radius has increased to 35.1 Å, with a shell thickness of 35 Å. A good fit is obtained with a hard-sphere volume fraction of 8.9%, which is again higher than the nominal $\Phi_{mic} = 7.1\%$, but less than in the oil-free case. This may indicate changes in interaction upon incorporation of polar oil. The hard-sphere radius of 60 Å is also in reasonable agreement with a dense core and a weak shell. At $c/c_{sat} = 50$, the core radius has increased to 41.2 Å, with still the same shell thickness of 35Å. A much weaker structure factor was used ($\Phi_{HS} = 1.5\%$, $R_{HS} = 70 \ \text{Å}$) leading to an underestimation of the peak. We have checked that using the nominal $\Phi$ value of 8.2% for $\Phi_{HS}$ gives a worsened but still acceptable description of the intensity at low-q, the best fit being obtained with the weaker S(q). To finish, the small quantity (10%) of PPO located in the shell necessary to describe the TBP-free data did not have any incidence on the fitting of the TBP-containing samples. This is due to the much stronger intensity from the more massive core, and we have thus set this value to zero.

**Structure in presence of TBP, after resolubilization :** At high oil concentrations, $c/c_{sat} >$ 150, resolubilisation is achieved, as it is shown in Fig.1a and b. The structure of this new microemulsion has been measured by SANS, and the cross sections are shown in Fig. 4. One immediately sees that the intensities are much higher, which indicates bigger objects. We have shifted the curves for convenience; without this shift, all three intensities would be very close, and in particular the Porod regime would overlap perfectly. The latter indicates that the specific surface S/V is the same for the three samples (A = 7.5 $10^{27}$ cm$^{-5}$, S/V = 3.21 $10^5$ cm$^{-1}$). The total micellized volume fractions of the samples c/c$_{sat}$ =190 (resp. 220 and 250) are $\Phi_{mic}$ = 14.4% (resp. 15.6% and 16.9%). As it was the case before the phase separation, this translates into now much higher Porod radii of 135 Å (resp. 146 and 159 Å). If we keep in mind the contribution of the PEO layer, our analysis is backed up by DLS results (Fig. 1a), where hydrodynamic radii of 160, 175, and 190 Å were found. From this size, a simple analysis assuming again a homogeneous distribution of copolymer and oil molecules in all micelles yields almost constant aggregation numbers of the order of 1000 for L64, and between 10000 and 18000 TBP molecules per micelle, increasing with oil concentration.

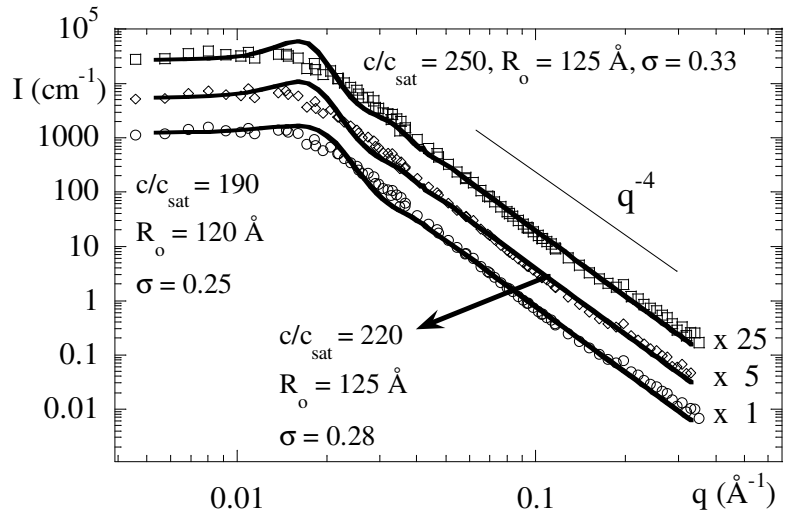

**Figure 4:** Scattered intensity of solutions of Pluronic L64 and TBP after resolubilization, at various concentrations in D$_2$O. The curves have been shifted by factors of 5 and 25. The mass fraction of Pluronics is fixed to 10%wt, with three concentrations of TBP (c/c$_{sat}$ = 190; 220; 250). The curves are compared to polydisperse spheres model calculations (**190**: N$_{L64}$ = 920, N$_{TBP}$ = 11800, R$_0$ = 120 Å, $\sigma$ = 25%; **220**: N$_{L64}$ = 1020, N$_{TBP}$ = 15200, R$_0$ = 125 Å, $\sigma$ = 28%; **250**: N$_{L64}$ = 1090, N$_{TBP}$ =18300 , R$_0$ = 125 Å, $\sigma$ = 33%) combined with a PY-structure factor (see text for details).

As before, we have proceeded to a complete fit of the experimental curves, in order to extract geometrical parameters. Core-shell modelling was found not to be necessary any more, and polydisperse spheres described the data well. This is reasonable, given that the already dominating core volume has increased much more than the one of the shell. The fits in Fig.4 have been obtained with a log-normal distribution ($R_o$, sigma) describing the core, the parameters being given in the caption. The typical core radii $R_o$ are approximately constant, whereas the polydispersity $\sigma$ - now much greater than before the phase transition - increases. Due to coupling of parameters, one could just as well have increased $R_o$ slightly, and kept $\sigma$ constant. The main point is that the volume per droplet increases, and the aggregation numbers deduced from these volumes are summarized in Table 1. For convenience, the dry micellar radius $R_{dry}$ (i.e. calculated from the L64 and TBP volumes only) is also reported. In all cases, the PY structure factor had to be used, and the data is again convoluted with the resolution function of the instrument. The structure factor parameters have been determined from the strong suppression in the low-q intensity, i.e. the ratio between the measured $I_0$ and the one expected in absence of interactions and estimated with the radius obtained from the Porod analysis. It turns out that the hard-sphere interaction is strong, and the volume fraction $\Phi_{HS}$ is higher than the actual volume fraction $\Phi_{mic}$. For the samples $c/c_{sat}$ =190 (resp. 220 and 250), we have obtained $\Phi_{HS}$ = 21% (resp. 27% and 28.9%). This corresponds to hard sphere radii $R_{HS}$ of 158 Å, 177 Å, and 188 Å, respectively, i.e. quite exactly the sum of the average radius (by volume, between 130 and 146 Å) and a constant of about 35±5 Å. Our natural interpretation is that the form factor of the core scattering is sufficient to describe the measured intensities, as its mass dominates over the hydrated shell, but the micellar interactions are mediated by the hydrodynamic interactions of the shells, of thickness 35 Å, as before the phase separation. The presence of the shell thus affects both the structure factor in SANS and the hydrodynamic radius measured by DLS, with good quantitative agreement.

| $c/c_{sat}$ | $\Phi_{mic}$(L64 and TBP) | N(L64) | N(TBP) | $R_{dry}$ (Å) | $R_o$(Å) | $\sigma$ | $\Delta$ (Å) |
|---|---|---|---|---|---|---|---|
| **0** | 6.0% | 35 | 0 | 33.8 | 27.9 | 15% | 35 |
| **25** | 7.1% | 49 | 80 | 39.9 | 35.1 | 15% | 35 |
| **50** | 8.2% | 65 | 233 | 46.1 | 41.2 | 15% | 35 |
| **190** | 14.4% | 920 | 11800 | 131.8 | 120.0 | 25% | 35 |
| **220** | 15.6% | 1020 | 15200 | 140.6 | 125.0 | 28% | 35 |
| **250** | 16.9% | 1090 | 18300 | 147.2 | 125.0 | 33% | 35 |

**Table 1 :** Characteristics of copolymer micelles (10%wt L64 in $D_2O$) swollen with TBP-oil. Geometrical quantities are defined in the appendix.

**Distribution of TBP inside droplets using contrast variation:** We have used isotopic substitution in order to change the neutron scattering contrasts of the two components, L64 and TBP, with respect to the solvent. This was achieved by using fully deuterated TBP (TBPd), and by mixing $H_2O$ and $D_2O$ (85%/15% by volume) such that the scattering length density of the L64 was matched by the solvent. The resulting scattering curves, which are shown in Figure 5, thus reflect the spatial distribution of the TBP molecules in the microemulsion droplets.

In Figure 5, the scattered intensities I(q) of 10%wt solutions of L64 at 308K containing different amounts of TBP are normalized by the contrast and volume fraction of matter contributing to the signal according to :

$$\frac{I(q)}{\Delta\rho^2\phi} = V_{dry} \cdot P(q) \cdot S(q) \qquad (3)$$

where P(q) and S(q) are the normalized form and structure factors, respectively. We have compared TBPd and fully hydrogenated TBP (TBPh) samples, and the scattering thus stems either only from the TBP (volume fraction 2.2%, contrast 5.34 $10^{10}$ cm$^{-2}$), or from both TBP and L64 (volume fraction 8.2%, contrast 6.1 $10^{10}$ cm$^{-2}$). This normalized intensity contains information on the structure factor, which is identical in the two cases, and on the form factor of the droplets. The changes in shape of the curves thus correspond to changes of droplet shape, and the scattering amplitude corresponds to the dry volume of scattering material per object, as described by the volume prefactor in eq.(3). As can be seen in the Figures, the behaviour is different before and after resolubilization.

Before emulsification failure, only the case c/c$_{sat}$ = 50 is shown in Figure 5 (cf. SI for c/c$_{sat}$ = 25). The low-q limit is obviously different, which reflects the different amounts of TBP and L64 present in each droplet. A quantitative analysis of the low-q intensities yields a mass (or volume) ratio of 4.3, which is compatible with the values given above (8.2%/2.2%). In the inset, we have superimposed the intensities normalized to one at low angles. It can be seen that the *shape* of the curve is indeed quite similar. However, the radius of gyration characterizing the spatial extend of either the complete droplet or the TBPd appears to be shifted to smaller radii in the case of TBPd. Its value is about 39 Å in the case of the complete droplet, but only 29 Å in the case where only the TBP is visible. One can conclude that the distribution of the TBP is homogeneous in the *core* of the droplets, and that the contribution of the hydrated shell is rather weak. Our findings thus suggest a droplet containing a homogenous mixture of TBP and PPO in the core, surrounded by a hydrated shell of PEO. The same conclusions hold for $c/c_{sat}$ = 25, and a sketch of the situation is shown in Figure 7.

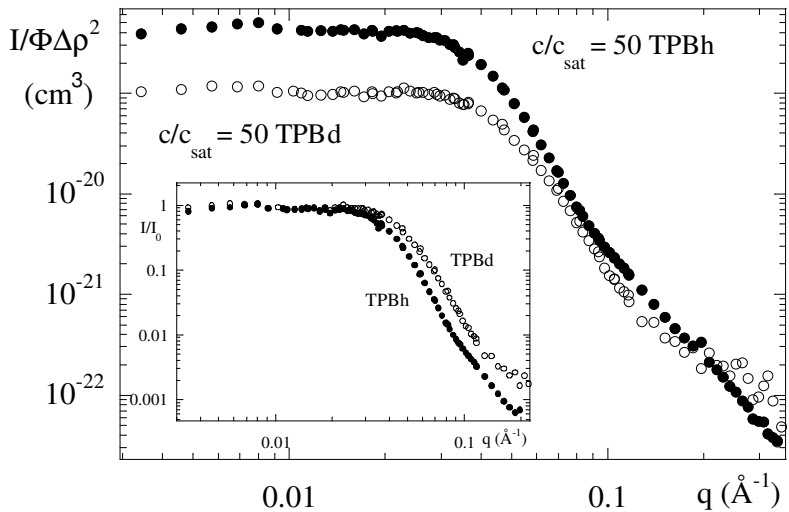

**Figure 5:** Renormalized intensity $I/\Phi\Delta\rho^2$ of solutions of Pluronic L64 (10%wt) and TBP before the phase separation, at $c/c_{sat}$ = 50. Two contrasts are compared (TBPd in $H_2O/D_2O$ contrast matching L64; L64 and TBPh in $D_2O$). $\Phi$ and $\Delta\rho$ refer to matter with contrast.

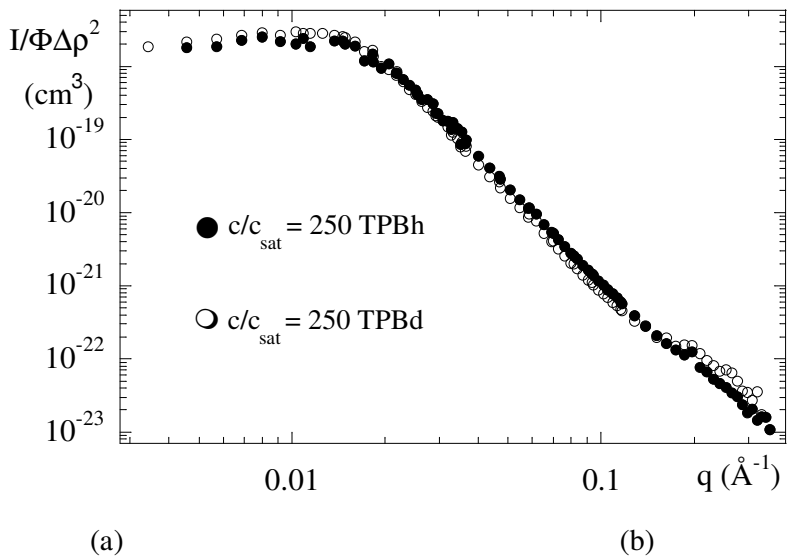

(a)                                        (b)

**Figure 6:** Renormalized intensity $I/\Phi\Delta\rho^2$ of solutions of Pluronic L64 (10%wt) and TBP after the phase separation, at $c/c_{sat}$ = 250. Two contrasts are compared (TBPd in $H_2O/D_2O$ contrast matching L64; L64 and TBPh in $D_2O$). $\Phi$ and $\Delta\rho$ refer to matter with contrast.

After resolubilization, we have performed a similar contrast variation experiment at $c/c_{sat} = 190$ and 250. Only the latter case is shown in Figure 6 (the other one being in the SI), where one can directly read from the normalized intensity that the structures are identical. This means that the contribution of the scattering of the copolymer, L64, is completely negligible with respect to the one of the TBP. This observation is in line with our conclusions from the modelling of the TBPh intensities, where we have found a large majority of TBP in each droplet. It also explains why the homogeneous sphere model is sufficient to describe the data, as the TBP making up the droplet core dominates the scattering. The situation is depicted in Figure 7. It represents an essentially pure TBP droplet core which is stabilized by a comparatively thin and hydrated PEO corona.

To summarize, the TBPd experiments thus confirm our modelling results before and after resolubilization. They further prove that TBP is intimately mixed to the copolymer in the case of the small droplets before emulsification failure ($c/c_{sat} < 70$), whereas it presents a pure nanophase stabilized by a shell of presumably hair-pin shaped L64 molecules after resolubization. In this context, it is interesting to calculate the average surface per PEO-group (either from the size distribution or directly from S/V): we find the surprising result that its value after resolubilization – 112 $\text{Å}^2$ – is only half of the one before the phase separation, 215 $\text{Å}^2$. For comparison, at the liquid-gaz interface a value of 169 $\text{Å}^2$ has been reported in the literature [85]. This gives additional credit to the idea that the hydrophilic groups have undergone reorganisation, now allowing the solubilization of much larger amounts of oil than before the emulsification failure. In term of local conformations, this finding is also in agreement with the observed decrease of the spontaneous curvature of the hydrophobic-hydrophilic interface.

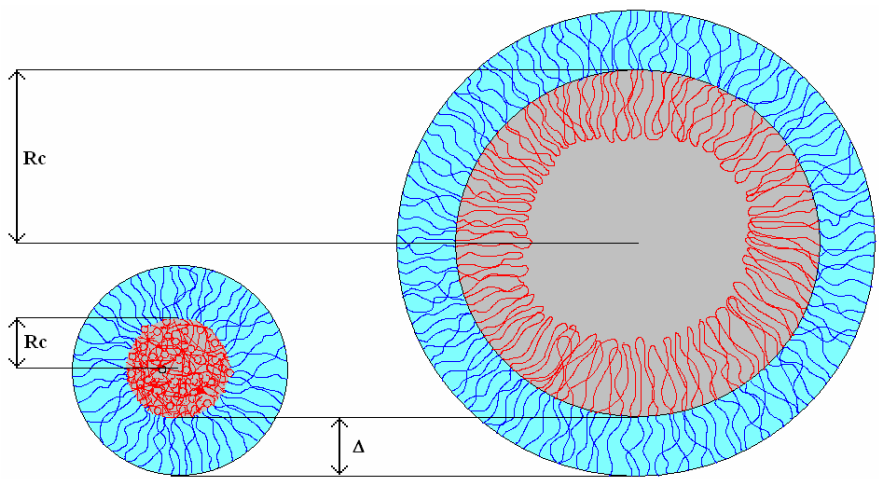

**Figure 7 :** Sketch of the TBP solubilisation in the L64 aggregates during the first (lowest TBP contents, i.e. $c/c_{sat} \leq 70$) and final domain of solubilisation (highest TBP concentrations, i.e. $c/c_{sat} \geq 150$).

## CONCLUDING REMARKS

The characterization of the micellar aggregates has been performed using SANS and extensive modelling, which included a core-shell description of the micelles with a coherent set of parameters, polydispersity, a hard-sphere structure factor, and the resolution function of the instrument. We have assumed that the almost water-insoluble oil TBP is inserted in the hydrophobic core, as confirmed by [1]H NMR analysis. Results for various Pluronics-systems proved that TBP molecules are preferably inserted into the hydrophobic core of the Pluronic micelles **[62]**. In the present SANS-analysis, it was found that small micelles are swollen with rather low amounts of TBP (some 10 % of micellar volume) before the emulsification failure (at $c/c_{sat} = 70$). The scattered intensities are correctly described by our model, which is based on the strong feature of a concentration-independent Porod domain at high q. At the other end of the q-scale, the low-q intensity suppression allowed the estimation of the hard sphere volume fraction needed for the PY calculation of S(q). $\Phi_{HS}$ is first found to be compatible with the core-shell picture, but with increasing oil content the structure factor is weakened, indicating less repulsive interactions. At higher oil content ($c/c_{sat} > 150$), a resolubilisation phenomenon occurs. The Porod domain is again found to be TBP-concentration independent, at a weaker level of specific surface compared to the low concentration case, which hints at a change in headgroup surface. We have performed the same type of analysis of the SANS patterns, and found considerably bigger emulsion droplets. It is interesting to note that the amount of incorporated TBP-oil increases dramatically with respect to the situation before the phase transition. On the other hand, the number of L64-molecules per micelle increases only slightly between $c/c_{sat} = 190$ and $c/c_{sat} = 250$, which is in line with a moderate increase of the average micellar surface. Last, but not least, our findings on the distribution of TBP and L64-copolymer are fully confirmed by an independent study using deuterated TBP.

To conclude, we have successfully described the structural changes underlying a rather surprising sequence of TBP-oil solubilization in a copolymer phase, followed by a phase separation, and resolubilization with increasing oil concentration. This structural investigation, using SANS and contrast variation, is hoped to contribute to ongoing solubilization studies of polar additives in microemulsions, with their possible applications to various types of decontamination.

## APPENDIX: SUMMARY OF THE SANS DATA ANALYSIS PROCEDURE

The SANS data has been analyzed in a step-by-step procedure. First, the unimer contribution was subtracted from the intensities, which modified only the pure L64 case, all others being dominated by the oil contribution. The free unimer concentration has been deduced from the difference between the experimentally observed low-q intensity (which is an average of micelles and unimers), and the prediction based on the literature value for the micellar aggregation number equal to 35. For a dispersion of identical, spherically symmetric micelles, the total scattered intensity can be written as:

$$I(Q) = \frac{N}{V} F_{mic}^2(q).S(q)$$
(A1)

where N/V is the number of micelles per unit volume, S(q) is the intermicellar structure factor, and $F_{mic}^2$ the square of the Fourier transform of the contrast, i.e. the (non normalized) form factor. For L64 in water, a form factor fit was performed at the nominal concentration of 5%, whereas a Percus-Yevick structure factor S(Q) had to be introduced for the 10% sample **[75, 76]**. Inspired by the amphiphilic nature of the block copolymer, a core-shell form factor was used to fit the data.

$$F^2_{mic.}(q, R_c) = \left[ V_c(\rho_c - \rho_s) f(qR_c) + V_s(\rho_s - \rho_{solv}) f(qR_s) \right]^2$$
(A2)

with

$$f(x) = 3 \frac{\sin(x) - x\cos(x)}{x^3}$$
$$R_s = R_c + \Delta$$
$$V_i = \frac{4\pi}{3} R_i^3$$
(A3)

Its parameters – core radius $R_c$, total radius including the shell $R_s$, shell thickness $\Delta$ and scattering length densities in both $\rho_s$, $\rho_c$ – can be constraint by a Guinier fit (eq. (1)). In the absence of intermicellar correlations, the low-q limit yields the total scattering length, which is usually transformed into mass or dry aggregate volume following eq.(3).

$$V_{dry} = \frac{I_0}{\Delta \rho^2 \Phi} = \frac{4\pi}{3} R_{dry}^3$$
(A4)

where $V_{dry}$ has been expressed through the dry radius $R_{dry}$. The second information from the Guinier fit is the overall radius of gyration (eq.(1)). This leaves us with only two parameters governing the distribution of matter between core and shell. A satisfying fit could be found assuming that the PPO made up the dry core ($\rho_c = \rho_{PPO}$), and PEO and solvent the shell according to $\rho_s = x.\rho_{PEO} + (1-x).\rho_{solv}$, with x the volume fraction of PEO chains in the shell. This fit could then be further improved by adding polydispersity in order to smooth the curves, as observed experimentally, allowing for some PPO-PEO exchange, and taking the resolution function into account. The polydispersity of the core can be described by means of a log-normal distribution with parameter $R_0$ and $\sigma$:

$$P(R, R_0, \sigma) = \frac{1}{\sqrt{2\pi}R\sigma} \exp\left(\frac{-1}{2\sigma^2} \ln^2 \frac{R}{R_0}\right)$$

(A5)

The form factor with polydispersity is calculated by integration over the distribution function:

$$F_{mic}^2(q) = \int P(R_c, R_0, \sigma).F_{mic}^2(q, R_c).dR_c$$

(A6)

For the 10% sample, we have first assumed that the form factor was the same as the one of the 5% sample. This leads to a self-consistent introduction of a structure factor, whose parameters are determined by the low-q intensity decrease as explained in the text.

The structure in presence of TBP has been analyzed assuming that all droplets are identical in composition. From the overlap in the Porod regime as TBP is added to a fixed amount of L64, it has been concluded that the copolymer decorates the interface. Equivalent Porod radii $R_{Porod}$ have been deduced from the specific surface S/V measured at large angles using eq.(2), and they have been used to estimate the number of TBP oil molecules and L64 in each droplet (cf. Table 1).

$$R_{Porod} = \frac{3\Phi}{S/V}$$

(A7)

More sophisticated core-shell fits of the oil-containing samples have been obtained following the same procedure as above, with the additional constraint of known mass replacing the low-q

Guinier information masked by the structure factor. Again, satisfying fits were obtained using a Percus-Yevick structure factor (which is constraint by the low-q isothermal compressibility), the resolution function, and polydispersity. In these cases, it proved unnecessary to allow a PEO-PPO exchange between core and shell, and the core characteristics are thus directly given by the mass and density of hydrophobic parts. Moreover, a description in terms of homogeneous spheres of oil-containing samples using the same formalism (eqs. (A1-A7)) was possible, with the form factor obtained from eq. (A2) in the limit of an invisible shell: $\rho_s = \rho_{solv.}$

A last technical point may be raised by close examination of our model fits. The agreement around the interaction peak is rather poor for the largest and most concentrated droplets. This may be due to the use of a monodisperse PY-structure factor, whereas the system polydispersity increases. It is well known, however, that polydispersity weakens the structure [86], and we have also checked that a slightly weaker structure factor can lead to very satisfying fits [87].